\documentclass[preprint,aps,showpacs,amsmath,amssymb]{revtex4}

\usepackage{graphics}% Include figure files
\usepackage{dcolumn}% Align table columns on decimal point
\usepackage{bm}% bold math

\begin{document}

\preprint{APS/123-QED}

\title{Algorithm for Generating Quasiperiodic Packings\\ of Multi-Shell Clusters}

\author{Nicolae Cotfas}
\email{ncotfas@yahoo.com}
\homepage{http://fpcm5.fizica.unibuc.ro/~ncotfas}
\affiliation{ Faculty of Physics, University of Bucharest,
PO Box 76--54, Bucharest 061513, Romania.}

\date{\today}% It is always \today, today,

\begin{abstract}
Many of the mathematical models used in quasicrystal physics are based
on tilings of the plane or space obtained by using
strip projection method in a superspace of dimension four, five or six. 
We present some mathematical results which allow one to use this very elegant method
in spaces of dimension much higher and to generate directly quasiperiodic packings
of multi-shell clusters. We show that in the case of a two-dimensional 
(resp. three-dimensional) cluster we have to compute only determinants of order 
three (resp. four), independently of the dimension of the superspace we use.
The computer program based on our mathematical results is very efficient.
For example, we can easily generate quasiperiodic packings of three-shell
icosahedral clusters (icosahedron + dodecahedron + icosidodecahedron)
by using strip projection method in a 31-dimensional space
(hundreds of points are obtained in a few minutes on a personal computer).
\end{abstract}

%Use showkeys class option if keyword
\maketitle

Quasicrystals are materials with perfect long-range order, 
but with no three-dimensional translational periodicity.
The discovery of these solids \cite{ds} in the early 1980's and the challenge 
to describe their structure led to a great interest in quasiperiodic 
sets of points. The diffraction image of a quasicrystal
often contains  a set of sharp Bragg peaks invariant under a finite
non-crystallographic group of symmetries $G$, called the symmetry group 
of quasicrystal (in reciprocal space). 
In the case of quasicrystals with no translational periodicity this group
is the icosahedral group $Y$ and in the case of quasicrystals 
periodic along one direction (two-dimensional quasicrystals) $G$ is one 
of the dihedral groups $D_8$ (octagonal quasicrystals), $D_{10}$ 
(decagonal quasicrystals) and $D_{12}$ (dodecagonal quasicrystals).

Real structure information obtained by high resolution transmission 
electron microscopy suggests us that a quasicrystal with symmetry group
$G$ can be regarded as a quasiperiodic packing of interpenetrating 
(partially occupied) copies of a well-defined $G$-invariant cluster $\mathcal{C}$. 
From a mathematical point of view, a $G$-cluster  is a finite union 
of orbits of $G$ in a fixed representation of $G$. A mathematical algorithm 
for generating quasiperiodic packings of interpenetrating copies of G-clusters
was proposed by author in collaboration with Jean-Louis Verger-Gaugry several
years ago \cite{jlvg,nc}. This algorithm based on strip projection method has been considered 
difficult to use since the dimension of the involved superspace is rather high.
The mathematical results we present in the present paper simplifies the 
computer program and allow to use strip projection method in superspaces
of large dimension. 

The dihedral group $D_{2m}$ can be defined in terms of generators and relations as
\begin{equation}
 D_{2m}=\langle \ a, \ b\ |\ \ a^{2m}=b^2=(ab)^2=e\ \rangle 
\end{equation}
and the relations
\begin{equation}
 a(\alpha ,\beta )=
\left(\alpha \, \cos \frac{\pi }{m}-\beta \, \sin \frac{\pi }{m}, \
      \alpha \, \sin \frac{\pi }{m}+\beta \, \cos \frac{\pi }{m} \right)
\qquad b(\alpha ,\beta )=(\alpha , -\beta ) 
\end{equation}
define an $\mathbb{R}$-irreducible representation in $\mathbb{R}^2$. 
The orbit of $D_{2m}$ generated by $(\alpha ,\beta )\in \mathbb{R}^2$
\begin{equation} D_{2m}(\alpha ,\beta )= \{ \ g(\alpha ,\beta )\ |\ g\in D_{2m}\ \}=
\{ (\alpha ,\beta ), a(\alpha ,\beta ),\, a^2(\alpha ,\beta ),\, ...,\, 
a^{2m-1}(\alpha ,\beta )\} \end{equation}
contains $2m$ points (vertices of a regular polygon with $2m$ sides). A two-shell
$D_{2m}$-cluster $\mathcal{C}_2$ is a union of two orbits
$\mathcal{C}_2=D_{2m}(\alpha _1,\beta _1) \cup D_{2m}(\alpha _2,\beta _2) .$

\begin{figure}
\includegraphics{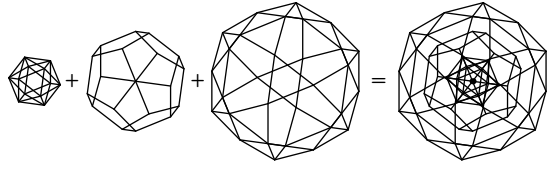}
\caption{A three-shell icosahedral cluster is a union of three orbits of $Y$.}
\end{figure}

The icosahedral group $Y=235$ can be defined in terms of generators and relations as 
\begin{equation} Y=\langle a,b\ |\ a^5=b^2=(ab)^3=e \rangle \end{equation}
and the rotations $a,\, b :\mathbb{R}^3\longrightarrow \mathbb{R}^3$
\begin{equation}\label{Y} \begin{array}{l}
a(\alpha ,\beta  ,\gamma )=
\left(\frac{\tau -1}{2}\alpha -\frac{\tau }{2}\beta  +\frac{1}{2}\gamma ,
\ \frac{\tau }{2}\alpha +\frac{1}{2}\beta  +\frac{\tau -1}{2}\gamma ,
\ -\frac{1}{2}\alpha +\frac{\tau -1}{2}\beta  
+\frac{\tau }{2}\gamma \right)\\[1mm]
b(\alpha ,\beta  ,\gamma )=(-\alpha ,-\beta  ,\gamma ).
\end{array} \end{equation}
where $\tau =(1+\sqrt{5})/2$, generate  an irreducible representation of $Y$ in 
$\mathbb{R}^3$.
In the case of this representation there are the trivial orbit 
$Y(0,0,0)=\{ (0,0,0)\}$ of length 1, the orbits
\begin{equation}
Y(\alpha ,\alpha \tau ,0)=\{ g(\alpha ,\alpha \tau ,0)\ |\ g\in Y\}\qquad 
{\rm where}\quad \alpha \in (0,\infty )
\end{equation}
of length 12 (vertices of a regular icosahedron), the orbits
\begin{equation}
Y(\alpha ,\alpha ,\alpha )=\{ g(\alpha ,\alpha ,\alpha )\ |\ g\in Y\}\qquad 
{\rm where}\quad \alpha \in (0,\infty )
\end{equation}
of length 20 (vertices of a regular dodecahedron), the orbits
\begin{equation}
Y(\alpha ,0,0)=\{ g(\alpha ,0,0)\ |\ g\in Y\}\qquad 
{\rm where}\quad \alpha \in (0,\infty )
\end{equation}
of length 30 (vertices of an icosidodecahedron), and all the other orbits are 
of length 60. The union of orbits $\mathcal{C}_3=
Y(\alpha ,\alpha \tau ,0)\cup Y(\beta , \beta ,\beta )\cup Y(\gamma , 0,0)\}$
is a three-shell icosahedral cluster (Fig. 1).

In order to have explicit mathematical formulae, we start by presenting our 
results in a particular case, namely, $G=D_{10}$. Let 
\begin{equation} \mathcal{C}_2=D_{10}(\alpha _1,\beta _1) \cup D_{10}(\alpha _2,\beta _2) 
=\{ v_1,\, v_2,\, ...,\, v_{10},\, -v_1,\, -v_2,\, ...,\, -v_{10} \} \end{equation}
where
\begin{equation} v_1=(v_{11},v_{21}),\quad  v_2=(v_{12},v_{22}),\quad ...,\quad 
v_{10}=(v_{1\, 10},v_{2\, 10})\end{equation}
be a fixed two-shell $D_{10}$-cluster, and let 
\begin{equation} w_1=(v_{11},v_{12},...,v_{1\, 10})\qquad \text{and}\qquad 
 w_2=(v_{21},v_{22},...,v_{2\, 10}).\end{equation}
From the general theory \cite{jlvg,nc} (a direct verification is also possible) it
follows that the vectors $w_1$ and $w_2$ from $\mathbb{R}^{10}$ have the same 
norm and are orthogonal
\begin{equation} \begin{array}{l}
v_{11}^2+v_{12}^2+...+v_{1\, 10}^2=
v_{21}^2+v_{22}^2+...+v_{2\, 10}^2\\ 
\langle w_1,w_2\rangle =v_{11}v_{21}+v_{12}v_{22}+...+v_{1\, 10}v_{2\, 10}=0 .
\end{array} \end{equation}

We identify the physical space with the two-dimensional subspace
\begin{equation} \bm{E}_2=\{ \ \alpha w_1+\beta w_2 \ | \ \alpha,\, \beta \in \mathbb{R}\ \} \end{equation}
of the superspace $\mathbb{R}^{10}$ and denote by $\bm{E}_2^\perp $ the 
orthogonal complement 
\begin{equation} \bm{E}_2^\perp =\{ \ x\in \mathbb{R}^{10}\ |\ 
\langle x,y\rangle =0\ \text{for\ all}\ y\in \bm{E}_2\ \}. \end{equation}
The orthogonal projection onto $\bm{E}_2$ of a vector $x\in \mathbb{R}^{10}$ is 
the vector
\begin{equation} \pi \, x= \left\langle x,\frac{w_1}{\kappa }\right\rangle\frac{w_1}{\kappa }+
             \left\langle x,\frac{w_2}{\kappa }\right\rangle\frac{w_2}{\kappa} \end{equation}
where $\kappa =||w_1||=||w_2||$, and the orthogonal projector corresponding 
to $\bm{E}_2^\perp $ is
\begin{equation} \pi ^\perp :\mathbb{R}^{10}\longrightarrow \bm{E}_2^\perp \qquad
\pi ^\perp x=x-\pi \, x. \end{equation}
If we describe $\bm{E}_2$ by using the orthogonal basis 
$\{ \kappa ^{-2}w_1,\, \kappa ^{-2}w_2\}$ then the orthogonal projector 
corresponding to $\bm{E}_2$ is
\begin{equation} \mathcal{P}_2: \mathbb{R}^{10}\longrightarrow \mathbb{R}^2\qquad 
\mathcal{P}_2x=(\langle x,w_1\rangle , \langle x,w_2\rangle ). \end{equation}

The projection $\bm{W}_{2,10}=\pi ^\perp (\bm{\Omega }_{10})$ of the unit hypercube
\begin{equation} \bm{\Omega }_{10}=[-0.5,\, 0.5]^{10}=\{ (x_1,x_2,...,x_{10})\ |\ -0.5\leq x_i\leq 0.5\
\text{for\ all\ } i\in \{ 1,2,..., 10\}\ \}. \end{equation}
is a polyhedron (called the {\it window} of selecton ) 
in the 8-dimensional subspace $\bm{E}_2^\perp $, 
and each 7-dimensional face of $\bm{W}_{2,10}$ is the projection of a 
7-dimensional face of $\bm{\Omega }_{10}$.
The vectors
\begin{equation} \begin{array}{rcl}
e_1&=&(1,0,0,0,0,0,0,0,0,0)\\
e_2&=&(0,1,0,0,0,0,0,0,0,0)\\
...&...&......................\\
e_{10}&=&(0,0,0,0,0,0,0,0,0,1)
\end{array} \end{equation}
form the canonical basis of $\mathbb{R}^{10}$, and each 7-face of $\bm{\Omega }_{10}$ 
is parallel to seven of these vectors and orthogonal to three of them.
There exist eight 7-faces of $\bm{\Omega }_{10}$ 
orthogonal to the distinct vectors $e_{i_1}$, $e_{i_2}$, $e_{i_3}$, and the set 
\begin{equation}\left\{ \ x=(x_1,x_2,...,x_{10})\ \left| \ \begin{array}{lcl}
x_i\in \{ -0.5,\, 0.5\} & \text{if}&   i\in \{ i_1,\, i_2,\, i_3\} \\
x_i=0 & \text{if}& i\not\in \{ i_1,\, i_2,\, i_3\}
\end{array} \right. \right\} \end{equation}
contains one and only one point from each of them.
There are 
\begin{equation} \left( \begin{array}{c}
10\\
3
\end{array}\right) =\frac{10\cdot 9\cdot 8}{1\cdot 2\cdot 3}=210 \end{equation}
sets of 8 parallel 7-faces of $\bm{\Omega }_{10}$, and we label them by using the 
elements of the set
\begin{equation} \mathcal{I}_{2,10}=\{ (i_1,i_2,i_3)\in \mathbb{Z}^3\ |\ 1\leq i_1\leq 8,\ \ i_1+1\leq i_2\leq 9,\ \ 
                                  i_2+1\leq i_3\leq 10\ \}. \end{equation}

In $\mathbb{R}^3$ the cross-product of two vectors $\bm{v}=(v_x,v_y,v_z)$ and
$\bm{w}=(w_x,w_y,w_z)$ is a vector orthogonal to $\bm{v}$ and $\bm{w}$, and
can be obtained by expanding the formal determinant
\begin{equation} \bm{v}\times \bm{w} = \left|
\begin{array}{ccc}
\bm{i} & \bm{j} & \bm{k} \\
v_x & v_y & v_z\\
w_x & w_y & w_z
\end{array} \right| \end{equation}
where $\{ \bm{i},\, \bm{j},\, \bm{k}\}$ is the canonical basis of $\mathbb{R}^3$. 
For any vector 
$\bm{u}=(u_x,u_y,u_z)$, the scalar product of $\bm{u}$ and $\bm{v}\times \bm{w}$ is 
\begin{equation} \bm{u}(\bm{v}\times \bm{w}) = \left|
\begin{array}{ccc}
u_x & u_y & u_z \\
v_x & v_y & v_z\\
w_x & w_y & w_z
\end{array} \right|. \end{equation}

In a very similar way, a vector $y$ orthogonal to nine vectors 
\begin{equation} u_i=(u_{i1},u_{i2},u_{i3},...,u_{i\, 10})\qquad i\in \{ 1,2,3,...,9\} \end{equation} 
from $\mathbb{R}^{10}$ can be obtained by expanding the formal determinant
\begin{equation} y=\left| \begin{array}{ccccc}
e_1 & e_2 & e_3 & ... & e_{10}\\
u_{11} & u_{12} & u_{13} & ... & u_{1\, 10}\\
u_{21} & u_{22} & u_{23} & ... & u_{2\, 10}\\
... & ... & ... & ... & ...\\
u_{91} & u_{92} & u_{93} & ... & u_{9\, 10}\
\end{array} \right| \end{equation}
containing the vectors of the canonical basis in the first row.
For any $x\in \mathbb{R}^{10}$, the scalar product of $x$ and $y$ is
\begin{equation} \langle x,y\rangle =\left| \begin{array}{ccccc}
x_1 & x_2 & x_3 & ... & x_{10}\\
u_{11} & u_{12} & u_{13} & ... & u_{1\, 10}\\
u_{21} & u_{22} & u_{23} & ... & u_{2\, 10}\\
... & ... & ... & ... & ...\\
u_{91} & u_{92} & u_{93} & ... & u_{9\, 10}\
\end{array} \right|. \end{equation}

For example, 
\begin{widetext}
\begin{equation} y=\left| \begin{array}{cccccccccc}
e_1 & e_2 & e_3 & e_4 & e_5 & 
e_6 & e_7 & e_8 & e_9 & e_{10}\\
0 & 0 & 0 & 1 & 0 & 0 & 0 & 0 & 0 & 0\\ 
0 & 0 & 0 & 0 & 1 & 0 & 0 & 0 & 0 & 0\\ 
0 & 0 & 0 & 0 & 0 & 1 & 0 & 0 & 0 & 0\\ 
0 & 0 & 0 & 0 & 0 & 0 & 1 & 0 & 0 & 0\\ 
0 & 0 & 0 & 0 & 0 & 0 & 0 & 1 & 0 & 0\\ 
0 & 0 & 0 & 0 & 0 & 0 & 0 & 0 & 1 & 0\\ 
0 & 0 & 0 & 0 & 0 & 0 & 0 & 0 & 0 & 1\\
v_{11} & v_{12} & v_{13} & v_{14} & v_{15} & 
v_{16} & v_{17} & v_{18} & v_{19} & v_{1\, 10}\\
v_{21} & v_{22} & v_{23} & v_{24} & v_{25} & 
v_{26} & v_{27} & v_{28} & v_{29} & v_{2\, 10}\\
\end{array} \right| 
= \left| \begin{array}{ccc}
e_1 & e_2 & e_3\\
v_{11} & v_{12} & v_{13}\\
v_{21} & v_{22} & v_{23}
\end{array} \right| \end{equation}
\end{widetext}
is a vector orthogonal to the vectors $e_4$, $e_5$, ...,  $e_{10}$, $w_1$, $w_2$, and
\begin{equation} \langle x,y\rangle =
\left| \begin{array}{ccc}
x_1& x_2 & x_3\\
v_{11} & v_{12} & v_{13}\\
v_{21} & v_{22} & v_{23}
\end{array} \right| \end{equation}
for any $x\in \mathbb{R}^{10}$. The vector  $y$ belongs to $\bm{E}_2^\perp $, and since
$e_i-\pi ^\perp e_i$ is a linear 
combination of $w_1$ and $w_2$, it is also orthogonal to $\pi ^\perp e_4$, 
$\pi ^\perp e_5$, ..., $\pi ^\perp e_{10}$. Therefore, $y$ is orthogonal 
to the 7-faces of $\bm{W}_{2,10}$ labelled by $(1,2,3)$. Similar results can be 
obtained for any $(i_1,i_2,i_3)\in \mathcal{I}_{2,10}$.

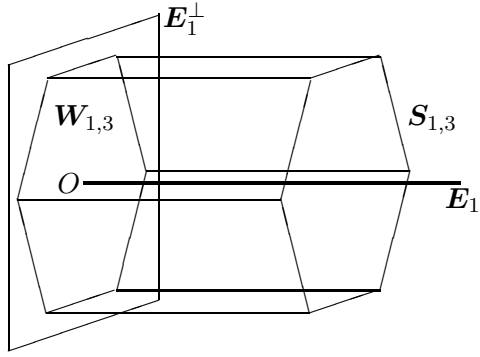
\begin{figure}
\setlength{\unitlength}{1mm}
\begin{picture}(70,50)(0,0)
\put(62,19){$\bm{E}_1$}
\put(24.5,43){$\bm{E}_1^\perp $}
\put(10.5,21){$O$}
\put(10,30){$\bm{W}_{1,3}$}
\put(57,30){$\bm{S}_{1,3}$}
\put(4,0){\line(0,1){38}}
\put(4,0){\line(3,1){20}}
\put(4,38){\line(3,1){20}}
\put(24,6.8){\line(0,1){38}}
\put(9,5){\line(3,1){9.1}}
\put(9,5){\line(1,0){35}}
\put(9,5){\line(-1,4){3.8}}
\put(5.2,20){\line(1,4){4}}
\put(5.2,20){\line(1,0){35}}
\put(18.4,8){\line(1,4){4}}
\put(18.4,8){\line(1,0){35}}
\put(22.3,23.9){\line(-1,4){3.8}}
\put(22.3,23.9){\line(1,0){35}}
\put(9.2,36.3){\line(3,1){9.1}}
\put(9.2,36.3){\line(1,0){35}}
\put(18.4,39.1){\line(1,0){35}}
\put(44,5){\line(3,1){9.1}}
\put(44,5){\line(-1,4){3.8}}
\put(40.2,20){\line(1,4){4}}
\put(53.4,8){\line(1,4){4}}
\put(57.3,23.9){\line(-1,4){3.8}}
\put(44.2,36.3){\line(3,1){9.1}}
\linethickness{0.4mm}
\put(14,22.3){\line(1,0){50}}
\end{picture}
\caption{The window $\bm{W}_{1,3}$ and the 
corresponding strip $\bm{S}_{1,3}$  
(case of a one-dimensional physical space 
$\bm{E}_1$ embedded into a three-dimensional superspace).}
\end{figure}

Consider the {\it strip} corresponding to $\bm{W}_{2,10}$ (Fig. 2)
\begin{equation} \bm{S}_{2,10}=\{ x\in \mathbb{R}^{10}\ |\ \pi ^\perp x\in \bm{W}_{2,10} \ \} \end{equation}
and define 
\begin{equation} d_{i_1i_2i_3}=\max_{\alpha _j \in \{ 0.5,\, 0.5\}}
\left| \begin{array}{ccc}
\alpha _1& \alpha _2 & \alpha _3 \\
v_{1i_1} & v_{1i_2} & v_{1i_3}\\
v_{2i_1} & v_{2i_2} & v_{2i_3}
\end{array} \right|\qquad \text{for\ each} \quad (i_1,i_2,i_3)\in \mathcal{I}_{2,10}.\end{equation}
A point $x=(x_1,x_2,...,x_{10})\in \mathbb{R}^{10}$ 
belongs to the strip $\bm{S}_{2,10}$ if and only if 
\begin{equation} -d_{i_1i_2i_3}\leq 
\left| \begin{array}{ccc}
x_{i_1}& x_{i_2} & x_{i_3}\\
v_{1i_1} & v_{1i_2} & v_{1i_3}\\
v_{2i_1} & v_{2i_2} & v_{2i_3}
\end{array} \right| 
\leq d_{i_1i_2i_3}\qquad \text{for\ any} \quad (i_1,i_2,i_3)\in \mathcal{I}_{2,10}.\end{equation}

The set defined in terms of the strip projection method \cite{ym,pk,md,ve,jhc}
\begin{equation} \mathcal{Q}_2=\mathcal{P}_2(\bm{S}_{2,10}\cap \mathbb{Z}^{10})=
\{ \ \mathcal{P}_2x\ | \ \ x\in \bm{S}_{2,10}\cap \mathbb{Z}^{10}\ \} \end{equation}
can be regarded (Fig. 3) as a quasiperiodic packing of translated (partially occupied)
copies of $\mathcal{C}_2$. Since
\begin{equation} \mathcal{P}_2e_i=(\langle e_i,w_1\rangle , \langle e_i, w_2\rangle )
=(v_{1i},v_{2i})=v_i \qquad
\text{for\ any\ }i\in \{ 1,2,...,10\}\end{equation} 
we get
\begin{eqnarray}
 \mathcal{P}_2&(\{ &x\pm e_1,\, x\pm e_2,\, ...,\, x\pm e_{10} \}\cap \bm{S}_{2,10})\nonumber \\
 &\subseteq &\{ \mathcal{P}_2x\pm v_1,\, \mathcal{P}_2x\pm v_2\, ...,\, 
\mathcal{P}_2x\pm v_{10}\}=\mathcal{P}_2x+\mathcal{C}_2 
\end{eqnarray}
that is, the neighbours of any point
$\mathcal{P}_2x\in \mathcal{Q}_2$ belong to the translated copy 
$\mathcal{P}_2x+\mathcal{C}_2$ of $\mathcal{C}_2 $.

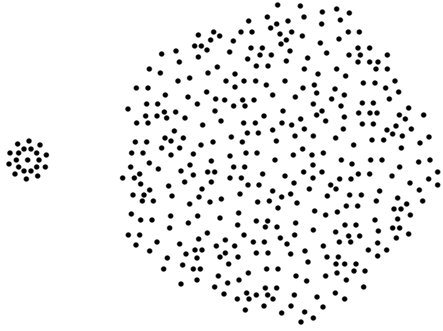
\begin{figure}
\setlength{\unitlength}{1.5mm}
\begin{picture}(30,40)(-20,-15)
\put(    .00000,    .00000){\circle*{0.5}} 
\put(  -1.10000,  -1.30000){\circle*{0.5}} 
\put(    .89645,  -1.44788){\circle*{0.5}} 
\put(   1.65404,    .40516){\circle*{0.5}} 
\put(    .12580,   1.69829){\circle*{0.5}} 
\put(  -1.57629,    .64444){\circle*{0.5}} 
\put(  -1.40902,  -2.25106){\circle*{0.5}} 
\put(   -.20355,  -2.74788){\circle*{0.5}} 
\put(  -2.67629,   -.65556){\circle*{0.5}} 
\put(   1.70547,  -2.03567){\circle*{0.5}} 
\put(   2.55049,  -1.04273){\circle*{0.5}} 
\put(   2.46306,    .99294){\circle*{0.5}} 
\put(   1.77984,   2.10344){\circle*{0.5}} 
\put(   -.18322,   2.64934){\circle*{0.5}} 
\put(  -1.45049,   2.34273){\circle*{0.5}} 
\put(  -2.57629,    .64444){\circle*{0.5}} 
\put(  -2.40902,  -2.25106){\circle*{0.5}} 
\put(   -.51256,  -3.69894){\circle*{0.5}} 
\put(  -2.98531,  -1.60662){\circle*{0.5}} 
\put(    .60547,  -3.33567){\circle*{0.5}} 
\put(  -3.67629,   -.65556){\circle*{0.5}} 
\put(   1.39645,  -2.98673){\circle*{0.5}} 
\put(   3.35951,  -1.63051){\circle*{0.5}} 
\put(   3.35951,   -.45494){\circle*{0.5}} 
\put(   3.27207,    .40516){\circle*{0.5}} 
\put(   2.58885,   2.69123){\circle*{0.5}} 
\put(   1.47082,   3.05450){\circle*{0.5}} 
\put(    .62580,   3.23713){\circle*{0.5}} 
\put(  -1.75951,   3.29378){\circle*{0.5}} 
\put(  -2.45049,   2.34273){\circle*{0.5}} 
\put(  -2.88531,   1.59550){\circle*{0.5}} 
\put(  -1.51256,  -3.69894){\circle*{0.5}} 
\put(  -3.98531,  -1.60662){\circle*{0.5}} 
\put(    .29645,  -4.28673){\circle*{0.5}} 
\put(   -.63836,  -5.39723){\circle*{0.5}} 
\put(  -3.98531,    .29550){\circle*{0.5}} 
\put(  -5.33033,  -1.06072){\circle*{0.5}} 
\put(   3.05049,  -2.58157){\circle*{0.5}} 
\put(   4.16853,  -1.04273){\circle*{0.5}} 
\put(   4.93580,  -2.27495){\circle*{0.5}} 
\put(   3.39787,   2.10344){\circle*{0.5}} 
\put(   2.27984,   3.64229){\circle*{0.5}} 
\put(   3.68885,   3.99123){\circle*{0.5}} 
\put(   -.95049,   3.88157){\circle*{0.5}} 
\put(  -2.75951,   3.29378){\circle*{0.5}} 
\put(  -2.65597,   4.74167){\circle*{0.5}} 
\put(  -1.63836,  -5.39723){\circle*{0.5}} 
\put(  -3.08885,  -3.05450){\circle*{0.5}} 
\put(  -5.63935,  -2.01177){\circle*{0.5}} 
\put(   1.95049,  -3.88157){\circle*{0.5}} 
\put(    .17066,  -5.98501){\circle*{0.5}} 
\put(  -5.63935,   -.10966){\circle*{0.5}} 
\put(  -3.85951,   1.99378){\circle*{0.5}} 
\put(   4.62679,  -3.22601){\circle*{0.5}} 
\put(   4.29433,    .65556){\circle*{0.5}} 
\put(   5.74482,  -1.68717){\circle*{0.5}} 
\put(   4.49787,   3.40344){\circle*{0.5}} 
\put(   3.37984,   4.94229){\circle*{0.5}} 
\put(    .70355,   4.28673){\circle*{0.5}} 
\put(  -1.84695,   5.32945){\circle*{0.5}} 
\put(  -3.65597,   4.74167){\circle*{0.5}} 
\put(   -.82934,  -5.98501){\circle*{0.5}} 
\put(  -2.73836,  -6.69723){\circle*{0.5}} 
\put(  -3.29240,  -5.80239){\circle*{0.5}} 
\put(  -3.21465,  -4.75279){\circle*{0.5}} 
\put(  -4.74289,  -3.45966){\circle*{0.5}} 
\put(  -6.44837,  -2.59956){\circle*{0.5}} 
\put(  -5.94837,  -1.06072){\circle*{0.5}} 
\put(  -6.73935,  -3.31177){\circle*{0.5}} 
\put(  -5.76515,  -3.71006){\circle*{0.5}} 
\put(  -7.21564,  -1.36733){\circle*{0.5}} 
\put(   1.82470,  -5.57985){\circle*{0.5}} 
\put(   3.52679,  -4.52601){\circle*{0.5}} 
\put(    .47967,  -6.93607){\circle*{0.5}} 
\put(   -.92934,  -7.28501){\circle*{0.5}} 
\put(   1.06711,  -7.43290){\circle*{0.5}} 
\put(   1.74695,  -6.62945){\circle*{0.5}} 
\put(  -6.53580,   1.33822){\circle*{0.5}} 
\put(  -5.51355,   1.58862){\circle*{0.5}} 
\put(  -7.21564,    .53478){\circle*{0.5}} 
\put(  -4.75597,   3.44167){\circle*{0.5}} 
\put(   5.43580,  -2.63822){\circle*{0.5}} 
\put(   5.52324,  -4.67389){\circle*{0.5}} 
\put(   4.50099,  -4.92429){\circle*{0.5}} 
\put(   5.39433,   1.95556){\circle*{0.5}} 
\put(   5.87062,    .01112){\circle*{0.5}} 
\put(   6.74482,  -1.68717){\circle*{0.5}} 
\put(   6.84482,   -.38717){\circle*{0.5}} 
\put(   6.64128,  -3.13505){\circle*{0.5}} 
\put(   7.39886,  -1.28201){\circle*{0.5}} 
\put(   4.18885,   4.35450){\circle*{0.5}} 
\put(   6.15191,   3.80860){\circle*{0.5}} 
\put(   6.07416,   2.75900){\circle*{0.5}} 
\put(   3.68885,   5.89334){\circle*{0.5}} 
\put(   2.48338,   6.39017){\circle*{0.5}} 
\put(   5.03388,   5.34744){\circle*{0.5}} 
\put(   3.50564,   6.64057){\circle*{0.5}} 
\put(   1.80355,   5.58673){\circle*{0.5}} 
\put(   -.19291,   5.73461){\circle*{0.5}} 
\put(  -2.84695,   5.32945){\circle*{0.5}} 
\put(   -.74695,   6.62945){\circle*{0.5}} 
\put(  -1.72115,   7.02774){\circle*{0.5}} 
\put(  -4.46498,   5.32945){\circle*{0.5}} 
\put(  -5.31001,   4.33651){\circle*{0.5}} 
\put(  -3.53017,   6.43995){\circle*{0.5}} 
\put(  -5.23226,   5.38611){\circle*{0.5}} 
\put(  -1.92934,  -7.28501){\circle*{0.5}} 
\put(  -4.39240,  -7.10239){\circle*{0.5}} 
\put(  -4.86869,  -5.15795){\circle*{0.5}} 
\put(  -7.54837,  -3.89956){\circle*{0.5}} 
\put(  -8.02466,  -1.95512){\circle*{0.5}} 
\put(  -7.52466,   -.41628){\circle*{0.5}} 
\put(  -6.86515,  -5.01006){\circle*{0.5}} 
\put(   3.40099,  -6.22429){\circle*{0.5}} 
\put(   -.62033,  -8.23607){\circle*{0.5}} 
\put(   1.37613,  -8.38395){\circle*{0.5}} 
\put(   2.64340,  -8.07734){\circle*{0.5}} 
\put(  -6.41001,   3.03651){\circle*{0.5}} 
\put(  -8.11210,   1.98266){\circle*{0.5}} 
\put(   6.33226,  -4.08611){\circle*{0.5}} 
\put(   5.39744,  -6.37218){\circle*{0.5}} 
\put(   6.97062,   1.31112){\circle*{0.5}} 
\put(   7.64128,  -3.13505){\circle*{0.5}} 
\put(   8.39886,  -1.28201){\circle*{0.5}} 
\put(   8.49886,    .01799){\circle*{0.5}} 
\put(   5.84289,   4.75966){\circle*{0.5}} 
\put(   7.72820,   3.16416){\circle*{0.5}} 
\put(   5.34289,   6.29850){\circle*{0.5}} 
\put(   3.81465,   7.59163){\circle*{0.5}} 
\put(   2.60918,   8.08846){\circle*{0.5}} 
\put(    .90709,   7.03461){\circle*{0.5}} 
\put(  -2.72115,   7.02774){\circle*{0.5}} 
\put(   -.62115,   8.32774){\circle*{0.5}} 
\put(  -4.33919,   7.02774){\circle*{0.5}} 
\put(  -6.04128,   5.97389){\circle*{0.5}} 
\put(  -6.88630,   4.98095){\circle*{0.5}} 
\put(  -1.62033,  -8.23607){\circle*{0.5}} 
\put(  -3.58338,  -7.69017){\circle*{0.5}} 
\put(  -5.20142,  -7.69017){\circle*{0.5}} 
\put(  -5.96869,  -6.45795){\circle*{0.5}} 
\put(  -7.67416,  -5.59785){\circle*{0.5}} 
\put(  -9.12466,  -3.25512){\circle*{0.5}} 
\put(  -8.33368,  -1.00406){\circle*{0.5}} 
\put(  -8.42111,   1.03161){\circle*{0.5}} 
\put(   4.29744,  -7.67218){\circle*{0.5}} 
\put(    .27613,  -9.68395){\circle*{0.5}} 
\put(   2.95242,  -9.02839){\circle*{0.5}} 
\put(  -7.98630,   3.68095){\circle*{0.5}} 
\put(  -8.92111,   2.57045){\circle*{0.5}} 
\put(   7.33226,  -4.08611){\circle*{0.5}} 
\put(   6.20646,  -5.78439){\circle*{0.5}} 
\put(   5.70646,  -7.32324){\circle*{0.5}} 
\put(   8.62466,   1.71628){\circle*{0.5}} 
\put(   9.29531,  -2.72989){\circle*{0.5}} 
\put(   9.49886,    .01799){\circle*{0.5}} 
\put(   6.15191,   5.71072){\circle*{0.5}} 
\put(   7.41919,   4.11522){\circle*{0.5}} 
\put(   8.72820,   3.16416){\circle*{0.5}} 
\put(   5.46869,   7.99679){\circle*{0.5}} 
\put(   2.91820,   9.03951){\circle*{0.5}} 
\put(   1.03289,   8.73290){\circle*{0.5}} 
\put(  -3.53017,   7.61552){\circle*{0.5}} 
\put(  -1.62115,   8.32774){\circle*{0.5}} 
\put(   -.31213,   9.27879){\circle*{0.5}} 
\put(  -5.91548,   7.67218){\circle*{0.5}} 
\put(  -7.69532,   5.56873){\circle*{0.5}} 
\put(   -.72387,  -9.68395){\circle*{0.5}} 
\put(  -3.27437,  -8.64123){\circle*{0.5}} 
\put(  -4.39240,  -8.27796){\circle*{0.5}} 
\put(  -4.89240,  -8.64123){\circle*{0.5}} 
\put(  -6.77771,  -7.04573){\circle*{0.5}} 
\put(  -9.32820,  -6.00300){\circle*{0.5}} 
\put(  -9.25046,  -4.95341){\circle*{0.5}} 
\put(  -9.43368,  -2.30406){\circle*{0.5}} 
\put( -10.77870,  -3.66028){\circle*{0.5}} 
\put(  -9.23013,    .44382){\circle*{0.5}} 
\put(  -9.23013,   1.61939){\circle*{0.5}} 
\put(   4.60646,  -8.62323){\circle*{0.5}} 
\put(    .15033, -11.38224){\circle*{0.5}} 
\put(   1.85242, -10.32839){\circle*{0.5}} 
\put(   2.82662, -10.72668){\circle*{0.5}} 
\put(  -8.79531,   4.26873){\circle*{0.5}} 
\put(  -9.73013,   1.98266){\circle*{0.5}} 
\put(   8.98630,  -3.68095){\circle*{0.5}} 
\put(   7.20646,  -5.78439){\circle*{0.5}} 
\put(   6.51548,  -6.73545){\circle*{0.5}} 
\put(   6.70646,  -7.32324){\circle*{0.5}} 
\put(   9.62466,   1.71628){\circle*{0.5}} 
\put(  10.39532,  -1.42989){\circle*{0.5}} 
\put(  10.87161,  -3.37433){\circle*{0.5}} 
\put(  11.07515,   -.62645){\circle*{0.5}} 
\put(   6.27771,   7.40900){\circle*{0.5}} 
\put(   7.72820,   5.06628){\circle*{0.5}} 
\put(   8.41919,   4.11522){\circle*{0.5}} 
\put(   9.03722,   4.11522){\circle*{0.5}} 
\put(   6.56869,   9.29679){\circle*{0.5}} 
\put(   4.57224,   9.44467){\circle*{0.5}} 
\put(   4.01820,  10.33951){\circle*{0.5}} 
\put(   1.34191,   9.68395){\circle*{0.5}} 
\put(  -2.43017,   8.91552){\circle*{0.5}} 
\put(  -5.10646,   8.25996){\circle*{0.5}} 
\put(  -1.31213,   9.27879){\circle*{0.5}} 
\put(  -1.12115,   9.86658){\circle*{0.5}} 
\put(  -6.81193,   9.12006){\circle*{0.5}} 
\put(  -7.56952,   7.26702){\circle*{0.5}} 
\put(  -8.59177,   7.01662){\circle*{0.5}} 
\put(  -2.37791, -10.08911){\circle*{0.5}} 
\put(   -.84967, -11.38224){\circle*{0.5}} 
\put(  -4.08338,  -9.22901){\circle*{0.5}} 
\put(  -5.01820, -10.33951){\circle*{0.5}} 
\put(  -6.46869,  -7.99679){\circle*{0.5}} 
\put(  -7.08673,  -7.99679){\circle*{0.5}} 
\put(  -7.87771,  -8.34573){\circle*{0.5}} 
\put(  -8.43175,  -7.45089){\circle*{0.5}} 
\put(  -6.90351,  -8.74402){\circle*{0.5}} 
\put( -10.42820,  -7.30300){\circle*{0.5}} 
\put( -10.90450,  -5.35856){\circle*{0.5}} 
\put( -10.33013,   -.85618){\circle*{0.5}} 
\put( -11.08772,  -2.70922){\circle*{0.5}} 
\put( -11.77870,  -3.66028){\circle*{0.5}} 
\put( -11.58772,  -3.07249){\circle*{0.5}} 
\put( -10.03915,   1.03161){\circle*{0.5}} 
\put(   5.60646,  -8.62323){\circle*{0.5}} 
\put(   5.41548,  -9.21102){\circle*{0.5}} 
\put(   5.50292, -10.07112){\circle*{0.5}} 
\put(   4.48066, -10.32152){\circle*{0.5}} 
\put(   6.18275,  -9.26768){\circle*{0.5}} 
\put(   -.15869, -12.33329){\circle*{0.5}} 
\put(   -.65869, -11.97003){\circle*{0.5}} 
\put(   1.72662, -12.02668){\circle*{0.5}} 
\put(   3.72308, -12.17456){\circle*{0.5}} 
\put(  -9.79531,   4.26873){\circle*{0.5}} 
\put(  -9.60433,   3.68095){\circle*{0.5}} 
\put(  -9.69177,   5.71662){\circle*{0.5}} 
\put( -10.44936,   3.86358){\circle*{0.5}} 
\put( -10.37161,   4.91317){\circle*{0.5}} 
\put( -11.38417,   1.57750){\circle*{0.5}} 
\put(   8.86050,  -5.37924){\circle*{0.5}} 
\put(  10.56259,  -4.32539){\circle*{0.5}} 
\put(   7.51548,  -6.73545){\circle*{0.5}} 
\put(   8.28275,  -7.96768){\circle*{0.5}} 
\put(   9.93367,   2.66733){\circle*{0.5}} 
\put(  10.43367,   2.30406){\circle*{0.5}} 
\put(  10.72466,   3.01628){\circle*{0.5}} 
\put(  11.27870,   2.12144){\circle*{0.5}} 
\put(  11.20095,   1.07184){\circle*{0.5}} 
\put(  11.97161,  -2.07433){\circle*{0.5}} 
\put(  11.68062,  -3.96212){\circle*{0.5}} 
\put(  11.18062,  -4.32539){\circle*{0.5}} 
\put(  12.72919,   -.22129){\circle*{0.5}} 
\put(   7.37771,   8.70900){\circle*{0.5}} 
\put(   7.85400,   6.76456){\circle*{0.5}} 
\put(   8.72820,   5.06628){\circle*{0.5}} 
\put(  10.13722,   5.41522){\circle*{0.5}} 
\put(   7.56869,   9.29679){\circle*{0.5}} 
\put(   7.37771,   9.88457){\circle*{0.5}} 
\put(   5.67224,  10.74467){\circle*{0.5}} 
\put(   4.14400,  12.03780){\circle*{0.5}} 
\put(   2.44191,  10.98395){\circle*{0.5}} 
\put(    .53289,  10.27174){\circle*{0.5}} 
\put(   1.03289,  10.63501){\circle*{0.5}} 
\put(    .44545,  11.13184){\circle*{0.5}} 
\put(   1.46770,  11.38224){\circle*{0.5}} 
\put(  -2.12115,   9.86658){\circle*{0.5}} 
\put(  -4.00646,   9.55996){\circle*{0.5}} 
\put(  -6.00292,   9.70785){\circle*{0.5}} 
\put(  -2.01761,  11.31446){\circle*{0.5}} 
\put(  -6.50292,  10.07112){\circle*{0.5}} 
\put(  -7.12095,  10.07112){\circle*{0.5}} 
\put(  -8.46597,   8.71490){\circle*{0.5}} 
\put( -10.16806,   7.66106){\circle*{0.5}} 
\put(  -3.18693, -10.67690){\circle*{0.5}} 
\put(  -2.50371, -11.78740){\circle*{0.5}} 
\put(  -1.65869, -11.97003){\circle*{0.5}} 
\put(  -4.20918, -10.92730){\circle*{0.5}} 
\put(  -5.32721, -11.29057){\circle*{0.5}} 
\put(  -6.59449,  -9.69507){\circle*{0.5}} 
\put(  -8.18673,  -9.29679){\circle*{0.5}} 
\put(  -7.21252,  -9.69507){\circle*{0.5}} 
\put(  -9.53175,  -8.75089){\circle*{0.5}} 
\put( -12.00450,  -6.65856){\circle*{0.5}} 
\put( -11.90450,  -5.35856){\circle*{0.5}} 
\put( -11.13915,   -.26839){\circle*{0.5}} 
\put( -11.98417,  -1.26134){\circle*{0.5}} 
\put( -11.89673,  -2.12144){\circle*{0.5}} 
\put( -12.58772,  -3.07249){\circle*{0.5}} 
\put( -12.48417,  -1.62461){\circle*{0.5}} 
\put( -11.69319,    .62645){\circle*{0.5}} 
\put(   7.18275,  -9.26768){\circle*{0.5}} 
\put(   6.31193, -10.65891){\circle*{0.5}} 
\put(   6.99177,  -9.85546){\circle*{0.5}} 
\put(   5.37712, -11.76941){\circle*{0.5}} 
\put(   -.96770, -12.92108){\circle*{0.5}} 
\put(   1.41761, -12.97774){\circle*{0.5}} 
\put(  -2.31273, -12.37518){\circle*{0.5}} 
\put(   2.62308, -13.47456){\circle*{0.5}} 
\put( -11.44936,   3.86358){\circle*{0.5}} 
\put( -11.37161,   4.91317){\circle*{0.5}} 
\put( -11.25837,   3.27579){\circle*{0.5}} 
\put( -11.26806,   6.36106){\circle*{0.5}} 
\put( -12.38417,   1.57750){\circle*{0.5}} 
\put(   9.16952,  -6.33029){\circle*{0.5}} 
\put(  10.43679,  -6.02368){\circle*{0.5}} 
\put(  10.87161,  -5.27645){\circle*{0.5}} 
\put(   9.09177,  -7.37989){\circle*{0.5}} 
\put(   9.09177,  -8.55546){\circle*{0.5}} 
\put(  11.03368,   3.96733){\circle*{0.5}} 
\put(  11.53368,   3.60406){\circle*{0.5}} 
\put(  12.08771,   2.70922){\circle*{0.5}} 
\put(  12.85499,   1.47700){\circle*{0.5}} 
\put(  12.78062,  -2.66212){\circle*{0.5}} 
\put(  13.62565,  -1.66917){\circle*{0.5}} 
\put(  11.98964,  -4.91317){\circle*{0.5}} 
\put(  11.05483,  -6.02368){\circle*{0.5}} 
\put(   8.37771,   8.70900){\circle*{0.5}} 
\put(   8.95400,   8.06456){\circle*{0.5}} 
\put(   8.85400,   6.76456){\circle*{0.5}} 
\put(   9.82820,   6.36628){\circle*{0.5}} 
\put(  10.94624,   6.00300){\circle*{0.5}} 
\put(   8.37771,   9.88457){\circle*{0.5}} 
\put(   9.14499,   8.65235){\circle*{0.5}} 
\put(   6.48126,  11.33246){\circle*{0.5}} 
\put(   5.79804,  12.44296){\circle*{0.5}} 
\put(   2.56770,  12.68224){\circle*{0.5}} 
\put(   -.36357,  11.71962){\circle*{0.5}} 
\put(    .13643,  12.08289){\circle*{0.5}} 
\put(   1.15869,  12.33330){\circle*{0.5}} 
\put(  -3.01761,  11.31446){\circle*{0.5}} 
\put(  -3.69744,  10.51102){\circle*{0.5}} 
\put(  -4.90292,  11.00785){\circle*{0.5}} 
\put(  -5.69390,  10.65891){\circle*{0.5}} 
\put(  -2.32662,  12.26552){\circle*{0.5}} 
\put(  -6.81193,  11.02218){\circle*{0.5}} 
\put(  -5.40292,  11.37112){\circle*{0.5}} 
\put(  -8.77499,   9.66596){\circle*{0.5}} 
\put( -10.04226,   9.35935){\circle*{0.5}} 
\put(  -3.31273, -12.37518){\circle*{0.5}} 
\put(  -4.51820, -11.87835){\circle*{0.5}} 
\put(  -6.90351, -10.64613){\circle*{0.5}} 
\put(  -9.84077,  -9.70195){\circle*{0.5}} 
\put(  -8.31252, -10.99507){\circle*{0.5}} 
\put( -11.10804,  -8.10645){\circle*{0.5}} 
\put( -13.00450,  -6.65856){\circle*{0.5}} 
\put( -12.31351,  -7.60962){\circle*{0.5}} 
\put( -12.71351,  -4.77078){\circle*{0.5}} 
\put( -12.79319,   -.67355){\circle*{0.5}} 
\put( -13.48417,  -1.62461){\circle*{0.5}} 
\put( -12.69319,    .62645){\circle*{0.5}} 
\put(   7.99177,  -9.85546){\circle*{0.5}} 
\put(   6.18613, -12.35719){\circle*{0.5}} 
\put(   7.88822, -11.30334){\circle*{0.5}} 
\put(   4.27712, -13.06940){\circle*{0.5}} 
\put(  -2.62174, -13.32624){\circle*{0.5}} 
\put(    .60859, -13.56552){\circle*{0.5}} 
\put(   2.31406, -14.42562){\circle*{0.5}} 
\put(   3.43209, -14.06235){\circle*{0.5}} 
\put( -12.25837,   3.27579){\circle*{0.5}} 
\put( -13.02565,   4.50802){\circle*{0.5}} 
\put( -12.26806,   6.36106){\circle*{0.5}} 
\put( -11.14226,   8.05935){\circle*{0.5}} 
\put(  10.74581,  -6.97473){\circle*{0.5}} 
\put(   9.90079,  -7.96768){\circle*{0.5}} 
\put(  11.84269,   4.55512){\circle*{0.5}} 
\put(  13.18771,   4.00922){\circle*{0.5}} 
\put(  13.66401,   2.06478){\circle*{0.5}} 
\put(  13.75145,    .02911){\circle*{0.5}} 
\put(  13.08964,  -3.61317){\circle*{0.5}} 
\put(  14.43466,  -2.25696){\circle*{0.5}} 
\put(  14.43466,  -1.08139){\circle*{0.5}} 
\put(  11.86384,  -6.61146){\circle*{0.5}} 
\put(   9.95400,   8.06456){\circle*{0.5}} 
\put(  10.63722,   6.95406){\circle*{0.5}} 
\put(   7.48126,  11.33246){\circle*{0.5}} 
\put(   9.95400,   9.24013){\circle*{0.5}} 
\put(   6.60705,  13.03074){\circle*{0.5}} 
\put(   5.48902,  13.39402){\circle*{0.5}} 
\put(   4.22174,  13.08740){\circle*{0.5}} 
\put(   2.25869,  13.63330){\circle*{0.5}} 
\put(   -.67258,  12.67068){\circle*{0.5}} 
\put(    .26223,  13.78118){\circle*{0.5}} 
\put( -20.91507,    .00000){\circle*{0.5}} 
\put( -22.91507,    .00000){\circle*{0.5}} 
\put( -21.60605,    .95106){\circle*{0.5}} 
\put( -22.22408,   -.95106){\circle*{0.5}} 
\put( -22.72408,    .58779){\circle*{0.5}} 
\put( -21.10605,   -.58779){\circle*{0.5}} 
\put( -22.72408,   -.58779){\circle*{0.5}} 
\put( -21.10605,    .58779){\circle*{0.5}} 
\put( -21.60605,   -.95106){\circle*{0.5}} 
\put( -22.22408,    .95106){\circle*{0.5}} 
\put( -20.81507,   1.30000){\circle*{0.5}} 
\put( -23.01507,  -1.30000){\circle*{0.5}} 
\put( -22.81152,   1.44788){\circle*{0.5}} 
\put( -21.01861,  -1.44788){\circle*{0.5}} 
\put( -23.56911,   -.40516){\circle*{0.5}} 
\put( -20.26103,    .40516){\circle*{0.5}} 
\put( -22.04087,  -1.69829){\circle*{0.5}} 
\put( -21.78927,   1.69829){\circle*{0.5}} 
\put( -20.33878,   -.64444){\circle*{0.5}} 
\put( -23.49136,    .64444){\circle*{0.5}} 
\put( -21.91507,    .00000){\circle*{0.5}}
\end{picture}
\caption{The two-shell decagonal cluster $\mathcal{C}_2$ and a fragment of
$\mathcal{Q}_2$.}
\end{figure}

Our algorithm works for any finite group $G$, any $\mathbb{R}$-irreducible 
representation of $G$ in a space $\mathbb{R}^n$ and any 
$G$-cluster symmetric with respect to the origin
\begin{equation} \mathcal{C}_n=
\{ v_1,\, v_2,\, ...,\, v_k,\, -v_1,\, -v_2,\, ...,\, -v_k \}\subset \mathbb{R}^n. 
\end{equation}
We identify the `physical space' with the $n$-dimensional  subspace
\begin{equation} \bm{E}_n=\{ \alpha _1w_1+\alpha _2w_2+...+\alpha _nw_n \ |\ \alpha _1,\, 
\alpha _2,\, ...,\, \alpha _n \in \mathbb{R} \ \} \end{equation}
of the superspace $\mathbb{R}^k$ spanned by the orthogonal vectors with the same norm
\begin{equation} w_i=(v_{i1},\ v_{i2},\, ...,\, v_{ik})\qquad i\in \{ 1,2 ,...,n\} \end{equation}
defined by using the coordinates $v_{ij}$ of the vectors 
$v_j=(v_{1j},v_{2j},...,v_{nj})$.

\begin{figure}			 
\setlength{\unitlength}{3mm}	  
\begin{picture}(20,25)(-20,-10) 
\put(   -3.01197,   -.12585){\circle*{0.2}} 
\put(   -3.90120,   2.26617){\circle*{0.2}} 
\put(   -4.24814,  -1.10535){\circle*{0.2}} 
\put(   -2.88674,  -3.12323){\circle*{0.2}} 
\put(   -1.69841,   -.99882){\circle*{0.2}} 
\put(   -2.32538,   2.33201){\circle*{0.2}} 
\put(   -3.56155,   1.35250){\circle*{0.2}} 
\put(   -2.46240,  -1.60419){\circle*{0.2}} 
\put(   -4.87511,   2.22548){\circle*{0.2}} 
\put(   -1.14884,  -2.47717){\circle*{0.2}} 
\put(   -5.56170,   -.23238){\circle*{0.2}} 
\put(    -.46225,   -.01931){\circle*{0.2}} 
\put(   -4.67248,  -2.62439){\circle*{0.2}} 
\put(   -1.35147,   2.37270){\circle*{0.2}} 
\put(   -3.43631,  -1.64488){\circle*{0.2}} 
\put(   -2.58764,   1.39319){\circle*{0.2}} 
\put(   -5.82396,  -1.17119){\circle*{0.2}} 
\put(    -.19999,    .91950){\circle*{0.2}} 
\put(   -5.01213,  -1.71072){\circle*{0.2}} 
\put(   -1.01182,   1.45903){\circle*{0.2}} 
\put(   -4.58779,   -.19169){\circle*{0.2}} 
\put(   -1.43616,   -.06000){\circle*{0.2}} 
\put(   -5.13737,   1.28666){\circle*{0.2}} 
\put(    -.88658,  -1.53835){\circle*{0.2}} 
\put(   -5.90136,    .68129){\circle*{0.2}} 
\put(    -.12259,   -.93298){\circle*{0.2}} 
\put(   -5.13736,   1.28666){\circle*{0.2}} 
\put(   -4.02644,   5.26356){\circle*{0.2}} 
\put(   -5.21476,   3.13915){\circle*{0.2}} 
\put(   -3.21461,   4.72403){\circle*{0.2}} 
\put(   -5.76434,   4.61749){\circle*{0.2}} 
\put(   -6.45093,   2.15964){\circle*{0.2}} 
\put(   -2.24070,   4.76472){\circle*{0.2}} 
\put(   -3.47686,   3.78521){\circle*{0.2}} 
\put(   -1.90105,   3.85105){\circle*{0.2}} 
\put(   -6.79058,   3.07331){\circle*{0.2}} 
\put(   -4.12290,  -4.10274){\circle*{0.2}} 
\put(   -5.56170,   -.23238){\circle*{0.2}} 
\put(   -4.93473,  -3.56321){\circle*{0.2}} 
\put(   -3.69856,  -2.58370){\circle*{0.2}} 
\put(   -6.79787,  -1.21189){\circle*{0.2}} 
\put(   -5.90864,  -3.60390){\circle*{0.2}} 
\put(   -3.82380,    .41368){\circle*{0.2}} 
\put(   -7.06012,  -2.15070){\circle*{0.2}} 
\put(   -2.67232,  -1.03951){\circle*{0.2}} 
\put(   -1.99751,  -5.51525){\circle*{0.2}} 
\put(   -1.57317,  -3.99621){\circle*{0.2}} 
\put(   -3.57333,  -5.58109){\circle*{0.2}} 
\put(   -2.33717,  -4.60158){\circle*{0.2}} 
\put(   -1.02360,  -5.47455){\circle*{0.2}} 
\put(   -4.54724,  -5.62178){\circle*{0.2}} 
\put(   -3.31107,  -4.64227){\circle*{0.2}} 
\put(   -4.88689,  -4.70811){\circle*{0.2}} 
\put(    -.76135,  -4.53574){\circle*{0.2}} 
\put(    -.80918,  -3.39084){\circle*{0.2}} 
\put(    -.46225,   -.01931){\circle*{0.2}} 
\put(   -1.01182,   1.45903){\circle*{0.2}} 
\put(   -2.24798,    .47952){\circle*{0.2}} 
\put(     .16473,  -3.35015){\circle*{0.2}} 
\put(     .85132,   -.89229){\circle*{0.2}} 
\put(   -2.12275,  -2.51786){\circle*{0.2}} 
\put(   -3.27423,  -1.06466){\circle*{0.2}} 
\put(    1.19097,  -1.80596){\circle*{0.2}} 
\put(   -1.08922,   3.31152){\circle*{0.2}} 
\put(   -2.45062,   5.32940){\circle*{0.2}} 
\put(   -2.87495,   3.81036){\circle*{0.2}} 
\put(   -4.18852,   4.68333){\circle*{0.2}} 
\put(     .22435,   2.43854){\circle*{0.2}} 
\put(    -.66488,   4.83056){\circle*{0.2}} 
\put(     .48660,   3.37736){\circle*{0.2}} 
\put(   -4.45077,   3.74452){\circle*{0.2}} 
\put(   -5.13737,   1.28666){\circle*{0.2}} 
\put(   -5.68694,   2.76501){\circle*{0.2}} 
\put(    -.59926,  -3.95552){\circle*{0.2}} 
\put(   -2.03806,   -.08515){\circle*{0.2}} 
\put(    -.88658,  -1.53835){\circle*{0.2}} 
\put(    -.33701,  -3.01670){\circle*{0.2}} 
\put(   -7.00050,   3.63799){\circle*{0.2}} 
\put(   -7.76449,   3.03262){\circle*{0.2}} 
\put(    1.40089,  -2.37064){\circle*{0.2}} 
\put(     .97655,  -3.88968){\circle*{0.2}} 
\put(    1.74055,  -3.28431){\circle*{0.2}} 
\put(   -8.37369,  -1.27773){\circle*{0.2}} 
\put(   -8.45109,    .57476){\circle*{0.2}} 
\put(    1.19826,   2.47923){\circle*{0.2}} 
\put(    2.34974,   1.02604){\circle*{0.2}} 
\put(    2.42714,   -.82645){\circle*{0.2}} 
\put(   -7.48446,  -3.66974){\circle*{0.2}} 
\put(   -6.67263,  -4.20927){\circle*{0.2}} 
\put(    -.92714,   3.89174){\circle*{0.2}} 
\put(    1.46051,   3.41805){\circle*{0.2}} 
\put(     .64868,   3.95758){\circle*{0.2}} 
\put(   -5.43647,  -3.22976){\circle*{0.2}} 
\put(   -5.01213,  -1.71072){\circle*{0.2}} 
\put(    -.58748,   2.97807){\circle*{0.2}} 
\put(   -1.01182,   1.45903){\circle*{0.2}} 
\put(   -7.82411,  -2.75607){\circle*{0.2}} 
\put(   -7.39978,  -1.23703){\circle*{0.2}} 
\put(   -7.94935,    .24131){\circle*{0.2}} 
\put(   -8.71334,   -.36406){\circle*{0.2}} 
\put(    1.80016,   2.50438){\circle*{0.2}} 
\put(    1.37583,    .98534){\circle*{0.2}} 
\put(    1.92540,   -.49300){\circle*{0.2}} 
\put(    2.68939,    .11237){\circle*{0.2}} 
\put(   -6.58795,  -1.77656){\circle*{0.2}} 
\put(   -7.90151,   -.90359){\circle*{0.2}} 
\put(     .56400,   1.52487){\circle*{0.2}} 
\put(    1.11357,    .04653){\circle*{0.2}} 
\put(    1.87756,    .65190){\circle*{0.2}} 
\put(   -6.71318,   1.22082){\circle*{0.2}} 
\put(     .68924,  -1.47251){\circle*{0.2}} 
\put(    1.45323,   -.86714){\circle*{0.2}} 
\put(   -8.02675,   2.09380){\circle*{0.2}} 
\put(    2.00280,  -2.34549){\circle*{0.2}} 
\put(   -6.45093,   2.15964){\circle*{0.2}} 
\put(   -7.68709,   1.18013){\circle*{0.2}} 
\put(   -4.71303,   2.80570){\circle*{0.2}} 
\put(   -3.33984,   7.72141){\circle*{0.2}} 
\put(   -5.88957,   7.61488){\circle*{0.2}} 
\put(   -2.36594,   7.76210){\circle*{0.2}} 
\put(   -3.60210,   6.78259){\circle*{0.2}} 
\put(   -7.07790,   5.49047){\circle*{0.2}} 
\put(   -7.76449,   3.03261){\circle*{0.2}} 
\put(   -4.79043,   4.65819){\circle*{0.2}} 
\put(   -5.07775,   7.07535){\circle*{0.2}} 
\put(   -1.55411,   7.22257){\circle*{0.2}} 
\put(   -5.34000,   6.13653){\circle*{0.2}} 
\put(   -7.88973,   6.03000){\circle*{0.2}} 
\put(   -8.65372,   5.42463){\circle*{0.2}} 
\put(   -9.34031,   2.96678){\circle*{0.2}} 
\put(   -1.81636,   6.28376){\circle*{0.2}} 
\put(    -.24054,   6.34960){\circle*{0.2}} 
\put(   -3.89992,   4.92324){\circle*{0.2}} 
\put(   -1.47671,   5.37009){\circle*{0.2}} 
\put(   -1.90105,   3.85105){\circle*{0.2}} 
\put(   -4.80949,  -6.56059){\circle*{0.2}} 
\put(   -3.57333,  -5.58109){\circle*{0.2}} 
\put(   -5.78340,  -6.60129){\circle*{0.2}} 
\put(   -5.01213,  -1.71072){\circle*{0.2}} 
\put(   -8.11143,   -.33891){\circle*{0.2}} 
\put(   -5.13736,   1.28666){\circle*{0.2}} 
\put(   -4.38516,  -5.04156){\circle*{0.2}} 
\put(   -7.48446,  -3.66974){\circle*{0.2}} 
\put(   -6.59523,  -6.06176){\circle*{0.2}} 
\put(   -7.74671,  -4.60856){\circle*{0.2}} 
\put(   -4.02522,  -3.75306){\circle*{0.2}} 
\put(   -6.24829,  -2.69023){\circle*{0.2}} 
\put(   -3.27423,  -1.06466){\circle*{0.2}} 
\put(   -2.12275,  -2.51786){\circle*{0.2}} 
\put(   -1.57317,  -3.99621){\circle*{0.2}} 
\put(   -8.45837,  -3.71043){\circle*{0.2}} 
\put(   -9.60985,  -2.25723){\circle*{0.2}} 
\put(   -9.68725,   -.40475){\circle*{0.2}} 
\put(   -8.72062,  -4.64925){\circle*{0.2}} 
\put(   -2.24798,    .47952){\circle*{0.2}} 
\put(    -.68395,  -6.38822){\circle*{0.2}} 
\put(    -.13438,  -7.86657){\circle*{0.2}} 
\put(   -3.65801,  -8.01379){\circle*{0.2}} 
\put(   -2.42185,  -7.03428){\circle*{0.2}} 
\put(   -3.99767,  -7.10012){\circle*{0.2}} 
\put(     .28996,  -6.34753){\circle*{0.2}} 
\put(   -1.99751,  -5.51525){\circle*{0.2}} 
\put(   -3.02376,  -7.05943){\circle*{0.2}} 
\put(   -1.71019,  -7.93241){\circle*{0.2}} 
\put(   -5.23383,  -8.07963){\circle*{0.2}} 
\put(    -.47403,  -6.95290){\circle*{0.2}} 
\put(    -.21177,  -6.01408){\circle*{0.2}} 
\put(    -.33907,  -7.31593){\circle*{0.2}} 
\put(   -2.68410,  -7.97310){\circle*{0.2}} 
\put(    1.10179,  -6.88706){\circle*{0.2}} 
\put(    1.86578,  -6.28169){\circle*{0.2}} 
\put(   -4.97158,  -7.14082){\circle*{0.2}} 
\put(   -7.35922,  -6.66713){\circle*{0.2}} 
\put(   -6.54740,  -7.20666){\circle*{0.2}} 
\put(   -5.31123,  -6.22715){\circle*{0.2}} 
\put(    1.05395,  -5.74216){\circle*{0.2}} 
\put(    1.74055,  -3.28431){\circle*{0.2}} 
\put(   -1.23352,  -4.90988){\circle*{0.2}} 
\put(     .22435,   2.43854){\circle*{0.2}} 
\put(   -1.01182,   1.45903){\circle*{0.2}} 
\put(    2.08748,    .08722){\circle*{0.2}} 
\put(    -.88658,  -1.53835){\circle*{0.2}} 
\put(   -2.03806,   -.08515){\circle*{0.2}} 
\put(   -1.56139,   2.93738){\circle*{0.2}} 
\put(    1.53791,   1.56557){\circle*{0.2}} 
\put(   -2.67232,  -1.03951){\circle*{0.2}} 
\put(   -3.82380,    .41368){\circle*{0.2}} 
\put(    2.71446,  -3.24361){\circle*{0.2}} 
\put(    3.05411,  -4.15728){\circle*{0.2}} 
\put(    2.06438,   -.84161){\circle*{0.2}} 
\put(     .42698,  -2.41133){\circle*{0.2}} 
\put(    3.66330,    .15306){\circle*{0.2}} 
\put(    3.74070,  -1.69943){\circle*{0.2}} 
\put(   -4.12290,  -4.10274){\circle*{0.2}} 
\put(   -3.69857,  -2.58370){\circle*{0.2}} 
\put(   -1.63879,   4.78987){\circle*{0.2}} 
\put(    1.46051,   3.41805){\circle*{0.2}} 
\put(     .57128,   5.81007){\circle*{0.2}} 
\put(   -3.00019,   6.80774){\circle*{0.2}} 
\put(   -4.31375,   7.68072){\circle*{0.2}} 
\put(    -.79012,   7.82794){\circle*{0.2}} 
\put(   -4.57601,   6.74190){\circle*{0.2}} 
\put(   -4.73809,   6.16168){\circle*{0.2}} 
\put(   -1.21445,   6.30890){\circle*{0.2}} 
\put(   -4.45077,   3.74452){\circle*{0.2}} 
\put(   -5.00035,   5.22286){\circle*{0.2}} 
\put(   -6.31391,   6.09584){\circle*{0.2}} 
\put(    1.88485,   4.93709){\circle*{0.2}} 
\put(    3.03633,   3.48389){\circle*{0.2}} 
\put(    -.13634,   6.72262){\circle*{0.2}} 
\put(    2.14710,   5.87591){\circle*{0.2}} 
\put(    1.33527,   6.41544){\circle*{0.2}} 
\put(   -5.56170,   -.23238){\circle*{0.2}} 
\put(   -7.13752,   -.29822){\circle*{0.2}} 
\put(   -7.26276,   2.69917){\circle*{0.2}} 
\put(   -7.55007,   5.11633){\circle*{0.2}} 
\put(   -8.57632,   3.57215){\circle*{0.2}} 
\put(    1.52613,  -5.36802){\circle*{0.2}} 
\put(    -.46225,   -.01931){\circle*{0.2}} 
\put(    1.11357,    .04653){\circle*{0.2}} 
\put(    1.23881,  -2.95086){\circle*{0.2}} 
\put(    2.55237,  -3.82384){\circle*{0.2}} 
\put(   -9.88988,   4.44512){\circle*{0.2}} 
\put(  -10.31422,   2.92608){\circle*{0.2}} 
\put(  -10.57648,   1.98727){\circle*{0.2}} 
\put(    4.29027,  -3.17777){\circle*{0.2}} 
\put(    2.55237,  -3.82384){\circle*{0.2}} 
\put(    3.86594,  -4.69681){\circle*{0.2}} 
\put(    4.55253,  -2.23896){\circle*{0.2}} 
\put(  -10.03419,  -3.77627){\circle*{0.2}} 
\put(  -10.37384,  -2.86260){\circle*{0.2}} 
\put(  -11.26307,   -.47059){\circle*{0.2}} 
\put(    4.01024,   3.52458){\circle*{0.2}} 
\put(    4.34989,   2.61091){\circle*{0.2}} 
\put(    5.23912,    .21890){\circle*{0.2}} 
\put(    4.55253,  -2.23896){\circle*{0.2}} 
\put(   -9.48461,  -5.25462){\circle*{0.2}} 
\put(  -10.37384,  -2.86260){\circle*{0.2}} 
\put(   -7.09697,  -5.72831){\circle*{0.2}} 
\put(   -8.24845,  -4.27511){\circle*{0.2}} 
\put(    1.07302,   5.47662){\circle*{0.2}} 
\put(    3.46067,   5.00293){\circle*{0.2}} 
\put(    4.34989,   2.61091){\circle*{0.2}} 
\put(    2.22450,   4.02342){\circle*{0.2}} 
\put(   -7.01229,  -3.29560){\circle*{0.2}} 
\put(    2.22450,   4.02342){\circle*{0.2}} 
\put(     .98834,   3.04391){\circle*{0.2}} 
\put(    1.11357,    .04653){\circle*{0.2}} 
\put(   -9.39993,  -2.82191){\circle*{0.2}} 
\put(  -10.71350,  -1.94894){\circle*{0.2}} 
\put(   -9.52517,    .17547){\circle*{0.2}} 
\put(  -10.83873,   1.04845){\circle*{0.2}} 
\put(    3.37598,   2.57022){\circle*{0.2}} 
\put(    3.92556,   1.09188){\circle*{0.2}} 
\put(    4.68955,   1.69725){\circle*{0.2}} 
\put(    3.50122,   -.42716){\circle*{0.2}} 
\put(    4.26521,    .17821){\circle*{0.2}} 
\put(    4.81478,  -1.30014){\circle*{0.2}} 
\put(   -8.71334,   -.36406){\circle*{0.2}} 
\put(   -9.47733,   -.96943){\circle*{0.2}} 
\put(  -10.02690,    .50892){\circle*{0.2}} 
\put(    2.68939,    .11237){\circle*{0.2}} 
\put(    3.45338,    .71774){\circle*{0.2}} 
\put(    4.00296,   -.76061){\circle*{0.2}} 
\put(   -9.60257,   2.02796){\circle*{0.2}} 
\put(    3.57862,  -2.27965){\circle*{0.2}} 
\put(   -9.00066,   2.05311){\circle*{0.2}} 
\put(   -6.02659,   3.67868){\circle*{0.2}} 
\put(  -10.57648,   1.98727){\circle*{0.2}} 
\put(   -5.13609,   3.94373){\circle*{0.2}} 
\put(   -4.16346,   1.32735){\circle*{0.2}} 
\put(   -3.13721,   2.87154){\circle*{0.2}} 
\put(   -5.20298,  10.07273){\circle*{0.2}} 
\put(   -1.67934,  10.21996){\circle*{0.2}} 
\put(   -4.22907,  10.11343){\circle*{0.2}} 
\put(   -5.46524,   9.13392){\circle*{0.2}} 
\put(   -8.01496,   9.02738){\circle*{0.2}} 
\put(   -8.77896,   8.42202){\circle*{0.2}} 
\put(   -1.94160,   9.28114){\circle*{0.2}} 
\put(   -4.02516,   7.92063){\circle*{0.2}} 
\put(   -9.62763,   5.38394){\circle*{0.2}} 
\put(   -6.65356,   7.00951){\circle*{0.2}} 
\put(   -9.96728,   6.29761){\circle*{0.2}} 
\put(   -7.34016,   4.55165){\circle*{0.2}} 
\put(  -10.65388,   3.83975){\circle*{0.2}} 
\put(   -5.21349,   5.79622){\circle*{0.2}} 
\put(   -7.20314,   8.48786){\circle*{0.2}} 
\put(   -1.02557,   9.11463){\circle*{0.2}} 
\put(   -1.12977,   8.74161){\circle*{0.2}} 
\put(     .44605,   8.80745){\circle*{0.2}} 
\put(   -5.76306,   7.27456){\circle*{0.2}} 
\put(   -8.22938,   6.94367){\circle*{0.2}} 
\put(  -10.77911,   6.83714){\circle*{0.2}} 
\put(   -9.07678,   6.56266){\circle*{0.2}} 
\put(  -11.20345,   5.31810){\circle*{0.2}} 
\put(  -12.15229,   1.92143){\circle*{0.2}} 
\put(   -2.23942,   7.42179){\circle*{0.2}} 
\put(     .18379,   7.86863){\circle*{0.2}} 
\put(    2.57144,   7.39494){\circle*{0.2}} 
\put(   -1.89977,   6.50812){\circle*{0.2}} 
\put(   -2.32411,   4.98908){\circle*{0.2}} 
\put(     .09911,   5.43593){\circle*{0.2}} 
\put(   -1.35147,   2.37270){\circle*{0.2}} 
\put(   -4.25992,  -8.03894){\circle*{0.2}} 
\put(   -6.46999,  -9.05914){\circle*{0.2}} 
\put(   -3.89999,  -6.75044){\circle*{0.2}} 
\put(   -1.71019,  -7.93241){\circle*{0.2}} 
\put(   -1.44794,  -6.99359){\circle*{0.2}} 
\put(   -8.59539,  -7.64663){\circle*{0.2}} 
\put(   -5.33878,  -2.88008){\circle*{0.2}} 
\put(   -7.56186,  -1.81726){\circle*{0.2}} 
\put(   -4.58779,   -.19168){\circle*{0.2}} 
\put(   -7.68709,   1.18013){\circle*{0.2}} 
\put(  -10.92341,  -1.38426){\circle*{0.2}} 
\put(  -11.00081,    .46823){\circle*{0.2}} 
\put(   -4.71181,  -6.21091){\circle*{0.2}} 
\put(   -6.93489,  -5.14809){\circle*{0.2}} 
\put(   -6.04566,  -7.54010){\circle*{0.2}} 
\put(   -9.14496,  -6.16829){\circle*{0.2}} 
\put(  -10.29644,  -4.71509){\circle*{0.2}} 
\put(   -9.40721,  -7.10710){\circle*{0.2}} 
\put(   -6.57495,  -3.85959){\circle*{0.2}} 
\put(   -3.60088,  -2.23402){\circle*{0.2}} 
\put(   -2.44940,  -3.68722){\circle*{0.2}} 
\put(   -1.89983,  -5.16556){\circle*{0.2}} 
\put(   -9.06028,  -3.73558){\circle*{0.2}} 
\put(   -1.69841,   -.99882){\circle*{0.2}} 
\put(     .00264,  -3.93037){\circle*{0.2}} 
\put(     .28996,  -6.34753){\circle*{0.2}} 
\put(  -11.27035,  -4.75578){\circle*{0.2}} 
\put(   -9.93650,  -3.42659){\circle*{0.2}} 
\put(  -12.49923,  -1.45010){\circle*{0.2}} 
\put(  -10.72078,  -6.23413){\circle*{0.2}} 
\put(    -.24783,   2.06440){\circle*{0.2}} 
\put(    1.17919,  -8.73955){\circle*{0.2}} 
\put(   -1.10828,  -7.90726){\circle*{0.2}} 
\put(     .55015,  -9.70794){\circle*{0.2}} 
\put(   -1.79488, -10.36512){\circle*{0.2}} 
\put(    -.55871,  -9.38561){\circle*{0.2}} 
\put(   -4.08235,  -9.53283){\circle*{0.2}} 
\put(   -5.65817,  -9.59867){\circle*{0.2}} 
\put(   -4.42200,  -8.61916){\circle*{0.2}} 
\put(     .97449,  -8.18890){\circle*{0.2}} 
\put(    2.83969,  -6.24100){\circle*{0.2}} 
\put(    3.17935,  -7.15467){\circle*{0.2}} 
\put(   -3.99767,  -7.10012){\circle*{0.2}} 
\put(   -1.16062,  -9.41076){\circle*{0.2}} 
\put(   -4.68426,  -9.55798){\circle*{0.2}} 
\put(   -1.02567,  -9.77378){\circle*{0.2}} 
\put(   -3.37070, -10.43096){\circle*{0.2}} 
\put(   -8.04581,  -9.12498){\circle*{0.2}} 
\put(   -7.23399,  -9.66451){\circle*{0.2}} 
\put(     .21050,  -8.79427){\circle*{0.2}} 
\put(    1.65136,  -8.36541){\circle*{0.2}} 
\put(    1.36404,  -5.94824){\circle*{0.2}} 
\put(   -1.99958,  -9.81448){\circle*{0.2}} 
\put(    1.78632,  -8.72843){\circle*{0.2}} 
\put(    2.55031,  -8.12306){\circle*{0.2}} 
\put(   -4.68426,  -9.55798){\circle*{0.2}} 
\put(    3.99117,  -7.69420){\circle*{0.2}} 
\put(    4.41551,  -6.17516){\circle*{0.2}} 
\put(   -6.97173,  -8.72569){\circle*{0.2}} 
\put(   -9.35938,  -8.25200){\circle*{0.2}} 
\put(   -5.86287,  -9.04803){\circle*{0.2}} 
\put(   -6.88705,  -6.29299){\circle*{0.2}} 
\put(    3.60368,  -5.63563){\circle*{0.2}} 
\put(     .62962,  -7.26120){\circle*{0.2}} 
\put(    2.95361,  -3.23362){\circle*{0.2}} 
\put(    1.31621,  -4.80334){\circle*{0.2}} 
\put(   -3.23368,  -6.49475){\circle*{0.2}} 
\put(   -2.80934,  -4.97572){\circle*{0.2}} 
\put(    -.32523,   3.91689){\circle*{0.2}} 
\put(    2.77407,   2.54507){\circle*{0.2}} 
\put(    1.53791,   1.56556){\circle*{0.2}} 
\put(   -1.43616,   -.06001){\circle*{0.2}} 
\put(   -2.58764,   1.39319){\circle*{0.2}} 
\put(    3.30055,    .13790){\circle*{0.2}} 
\put(    1.66315,  -1.43182){\circle*{0.2}} 
\put(   -2.46240,  -1.60419){\circle*{0.2}} 
\put(  -20.00000,    .00000){\circle*{0.2}} 
\put(  -20.00000,    .00000){\circle*{0.2}} 
\put(  -19.68834,   -.83837){\circle*{0.2}} 
\put(  -20.31166,    .83837){\circle*{0.2}} 
\put(  -19.25508,   -.49507){\circle*{0.2}} 
\put(  -20.74492,    .49507){\circle*{0.2}} 
\put(  -19.29897,    .55548){\circle*{0.2}} 
\put(  -20.70103,   -.55548){\circle*{0.2}} 
\put(  -19.75936,    .86145){\circle*{0.2}} 
\put(  -20.24064,   -.86145){\circle*{0.2}} 
\put(  -19.10635,    .03734){\circle*{0.2}} 
\put(  -20.89365,   -.03734){\circle*{0.2}} 
\put(  -19.04840,    .75403){\circle*{0.2}} 
\put(  -20.95160,   -.75403){\circle*{0.2}} 
\put(  -18.98882,   -.67202){\circle*{0.2}} 
\put(  -21.01118,    .67202){\circle*{0.2}} 
\put(  -18.36387,  -1.08735){\circle*{0.2}} 
\put(  -21.63613,   1.08735){\circle*{0.2}} 
\put(  -18.03722,    .08201){\circle*{0.2}} 
\put(  -21.96278,   -.08201){\circle*{0.2}} 
\put(  -18.46028,   1.22004){\circle*{0.2}} 
\put(  -21.53972,  -1.22004){\circle*{0.2}} 
\put(  -19.47146,   1.89206){\circle*{0.2}} 
\put(  -20.52854,  -1.89206){\circle*{0.2}} 
\put(  -20.42306,   1.13803){\circle*{0.2}} 
\put(  -19.57694,  -1.13803){\circle*{0.2}} 
\put(  -20.32665,  -1.16936){\circle*{0.2}} 
\put(  -19.67335,   1.16936){\circle*{0.2}} 
\put(  -19.31547,  -1.84137){\circle*{0.2}} 
\put(  -20.68453,   1.84137){\circle*{0.2}} 
\put(  -18.78693,    .05068){\circle*{0.2}} 
\put(  -21.21307,   -.05068){\circle*{0.2}} 
\put(  -20.88923,   2.39202){\circle*{0.2}} 
\put(  -19.11077,  -2.39202){\circle*{0.2}} 
\put(  -21.23616,   -.97951){\circle*{0.2}} 
\put(  -18.76384,    .97951){\circle*{0.2}} 
\put(  -19.87476,  -2.99739){\circle*{0.2}} 
\put(  -20.12524,   2.99739){\circle*{0.2}} 
\put(  -18.68644,   -.87298){\circle*{0.2}} 
\put(  -21.31356,    .87298){\circle*{0.2}} 
\put(  -19.31341,   2.45786){\circle*{0.2}} 
\put(  -20.68659,  -2.45786){\circle*{0.2}} 
\put(  -19.45043,  -1.47835){\circle*{0.2}} 
\put(  -20.54957,   1.47835){\circle*{0.2}} 
\put(  -18.13686,  -2.35132){\circle*{0.2}} 
\put(  -21.86314,   2.35132){\circle*{0.2}} 
\put(  -17.45027,    .10653){\circle*{0.2}} 
\put(  -22.54973,   -.10653){\circle*{0.2}} 
\put(  -18.33950,   2.49855){\circle*{0.2}} 
\put(  -21.66050,  -2.49855){\circle*{0.2}} 
\put(  -19.57566,   1.51904){\circle*{0.2}} 
\put(  -20.42434,  -1.51904){\circle*{0.2}} 
\put(  -17.18802,   1.04535){\circle*{0.2}} 
\put(  -22.81198,  -1.04535){\circle*{0.2}} 
\put(  -17.99984,   1.58488){\circle*{0.2}} 
\put(  -22.00016,  -1.58488){\circle*{0.2}} 
\put(  -18.42418,    .06584){\circle*{0.2}} 
\put(  -21.57582,   -.06584){\circle*{0.2}} 
\put(  -17.87461,  -1.41251){\circle*{0.2}} 
\put(  -22.12539,   1.41251){\circle*{0.2}} 
\put(  -17.11062,   -.80714){\circle*{0.2}} 
\put(  -22.88938,    .80714){\circle*{0.2}} 
\end{picture}		    
\caption{Projected positions of points of $\mathcal{C}_3$ and of $\mathcal{Q}_3$ 
down a fivefold axis in the case of an icosahedral three-shell cluster.}  
\end{figure}
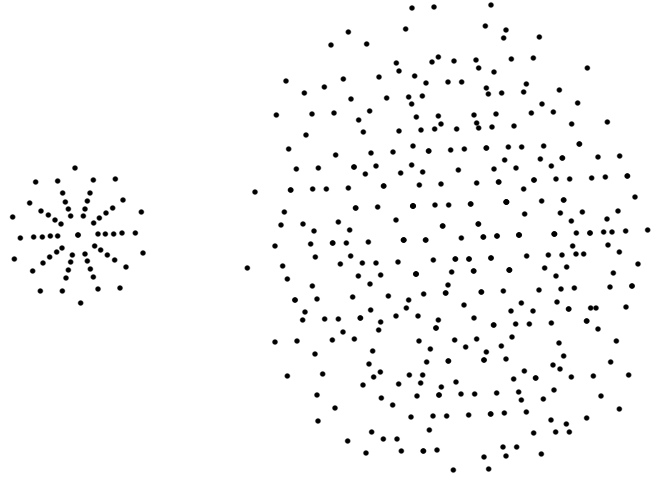

The window $\bm{W}_{n,k}=\pi ^\perp (\bm{\Omega }_k)$ is a polyhedron in the 
$(k-n)$-subspace $\bm{E}_n^\perp $. Each $(k-n-1)$-face of $\bm{W}_{n,k}$ is the
projection of a $(k-n-1)$-face of the unit hypercube $\bm{\Omega }_k=[-0.5,\, 0.5]^k$.
Each $(k-n-1)$-face of $\bm{\Omega }_k$ is parallel to $k-n-1$ vectors of the canonical
basis $\{ e_1,\, e_2,\, ...,\, e_k\}$ and orthogonal to $n+1$ of them.
For each $n+1$ distinct vectors $e_{i_1}$, $e_{i_2}$, ..., $e_{i_{n+1}}$ the number
of $(k-n-1)$-faces of $\bm{\Omega }_k$ orthogonal to them is $2^{n+1}$.
There are $k!/[(n+1)!\, (k-n-1)!]$ sets of $2^{n+1}$ parallel $(k-n-1)$-faces of $\bm{\Omega }_k$.
In the case $n=2$ these sets are labelled by 
\begin{equation} \mathcal{I}_{2,k}=
\{ (i_1,i_2,i_3)\in \mathbb{Z}^3\ |\ 1\leq i_1\leq k-2,\ \ i_1+1\leq i_2\leq k-1,\ \ 
                                  i_2+1\leq i_3\leq k\ \} \end{equation}
and in the case $n=3$ are labelled by
\begin{equation} \mathcal{I}_{3,k}=
\left\{ (i_1,i_2,i_3,i_4)\in \mathbb{Z}^4\ \left|\ 
            \begin{array}{rl}
            1\leq i_1\leq k-3,\ \ \ & i_1+1\leq i_2\leq k-2,\\ 
            i_2+1\leq i_3\leq k-1,\ \ \  & i_3+1\leq i_4\leq k
            \end{array} \right.
\right\}. \end{equation}
 In the case $n=3$, a point $x=(x_1,x_2,...,x_k)\in \mathbb{R}^k$ belongs to the
strip $\bm{S}_{3,k}$ if and only if 
\begin{equation} -d_{i_1i_2i_3i_4}\leq 
\left| \begin{array}{cccc}
x_{i_1}& x_{i_2} & x_{i_3} & x_{i_4}\\
v_{1i_1} & v_{1i_2} & v_{1i_3} & v_{1i_4}\\
v_{2i_1} & v_{2i_2} & v_{2i_3} & v_{2i_4}\\
v_{3i_1} & v_{3i_2} & v_{3i_3} & v_{3i_4}
\end{array} \right| 
\leq d_{i_1i_2i_3i_4}\qquad \text{for\ each\ } (i_1,i_2,i_3i_4)\in \mathcal{I}_{3,k}
\end{equation}
where
\begin{equation} d_{i_1i_2i_3i_4}=\max_{\alpha _j \in \{ 0.5,\, 0.5\}}
\left| \begin{array}{cccc}
\alpha _1& \alpha _2 & \alpha _3 & \alpha _4 \\
v_{1i_1} & v_{1i_2} & v_{1i_3} & v_{1i_4}\\
v_{2i_1} & v_{2i_2} & v_{2i_3} & v_{2i_4}\\
v_{3i_1} & v_{3i_2} & v_{3i_3} & v_{3i_4}
\end{array} \right|.\end{equation}

The pattern defined in terms of the strip projection method 
\begin{equation} \mathcal{Q}_n=\mathcal{P}_n(\bm{S}_{n,k}\cap \mathbb{Z}^k)=
\{ \mathcal{P}_nx\ |\ \ x\in \bm{S}_{n,k}\cap \mathbb{Z}^k\ \} \end{equation}
where 
\begin{equation} \mathcal{P}_n: \mathbb{R}^k\longrightarrow \mathbb{R}^n\qquad 
\mathcal{P}_nx=(\langle x,w_1\rangle , \langle x,w_2\rangle,..., 
 \langle x,w_n\rangle ), \end{equation}
can be regarded as a quasiperiodic packing of copies of the starting cluster 
$\mathcal{C}_n$.

Projected positions down a fivefold axis of points of a fragment of the 
pattern $\mathcal{Q}_3$ 
obtained by starting from the icosahedral three-shell cluster 
\begin{equation} \mathcal{C}_3=Y\left(\frac{1}{\sqrt{2+\tau }},
 \frac{\tau }{\sqrt{2+\tau }}, 0\right)\cup
   Y\left( \frac{2}{\sqrt{3}},\frac{2}{\sqrt{3}},
\frac{2}{\sqrt{3}}\right)\cup Y(3,0,0)\end{equation}
are presented in Fig. 4. The dimension $k$ of the superspace in this case is 31.

The description of the atomic structure of quasicrystals is a very difficult problem. 
Elser \& Henley \cite{eh} and  Audier \& Guyot \cite{ag} 
have obtained  models for icosahedral quasicrystals 
by decorating the Ammann rhombohedra occuring in a tiling of the 3D space
defined by projection. In his quasi-unit cell picture Steinhardt 
\cite{st} has shown that the atomic 
structure can be described entirely by using a single repeating cluster which overlaps
(shares atoms with) neighbour clusters. The model is determined by the overlap 
rules and the atom decoration of the unit cell. Some important models have been
obtained by Yamamoto \& Hiraga \cite{yh,ya}, 
Katz \& Gratias \cite{kg}, Gratias Puyraimond and Quiquandon \cite{gpq} 
by using the section method in a
six-dimensional superspace decorated with several polyhedra (acceptance domains).
Janot and de Boissieu \cite{jb} have shown that a model of icosahedral quasicrystal
can be generated recursively by starting from a pseudo-Mackay cluster 
and using some inflation rules. In the case of all these models one has to
add or shift some points in order to fill the gaps between the clusters, 
and one has to elliminate some points from interpenetrating clusters if they
become too close. 

The strip projection method has been used (to our knowlege) mainly for generating
tilings of plane or space, and only in superspaces of dimension four, five or six.
Algorithms and details concerning computer programs for generating quasiperiodic tilings 
have been presented by Conway and Knowles \cite{jhc}, 
Vogg and Ryder \cite{vr},Lord, Ramakrishnan and Ranganathan \cite{lrr}. 
We present an algorithm for generating quasiperiodic packings of multi-shell
clusters based on strip projection method used in a superspace of large dimension.
The quasiperiodic sets obtained in this way have the remarkable mathematical
properties of the patterns obtained by projection  and the desired local structure.
Each point of the set is the centre of a more or less occupied copy of the 
starting cluster, and the clusters corresponding to neighbouring points share 
several points.  Computer programs in FORTRAN 90 are available via internet \cite{cp}.

\end{document}